\documentstyle[11pt]{article}
\title{\vspace{-1.in} \hfill {\small\rm NTUA-100/00} \\~\\~\\
 Mass generation and the dynamical role \\ of the Katoptron Group}
\author{ George
Triantaphyllou\thanks{e-mail:gtriant$@$telesis.gr}
\\{\it Physics Department, National Technical University,
Athens}\\ {\it Zografou Campus, GR-157 80 Athens, GREECE }}
\begin{document}
\setlength{\baselineskip}{24pt}
\maketitle
\thispagestyle{empty}

\begin{abstract}
Heavy mirror fermions along with a new strong gauge interaction
capable of breaking the electroweak gauge symmetry dynamically
were recently introduced under the name of katoptrons. Their main
function is to provide a viable alternative to the Standard-Model
Higgs sector. In such a framework, ordinary fermions acquire
masses after the breaking of the strong katoptron group which
allows mixing with their katoptron partners. The purpose of this
paper is to study the elementary-scalars-free mechanism
responsible for this breaking and its implications for the fermion
mass hierarchies.
 \vspace{2.in}
\end{abstract}
\vfill
\setcounter{page}{0}
\pagebreak
\section{Motivation}

Identifying the nature of the various symmetries which might be at
the source of the observed fermion-mass pattern constitutes one of
the central investigation goals in high-energy physics
\footnote{``Of symmetries indeed, we consider the small which we
perceive, we neglect however the principal and greatest" wrote
Plato in his dialog ``Timaios".}.
 The breaking of an horizontal
symmetry for instance was used some time ago in an attempt to
explain the mass hierarchy between fermion generations
\cite{Zoup}. The katoptron theory that was recently introduced in
\cite{george1} contains a strongly interacting ``horizontal" gauge
symmetry acting on a new fermion sector. It constitutes a
dynamical alternative to the Standard-Model (SM) Higgs mechanism
and in parallel addresses the strong CP problem. Strong dynamics
render the study of such theories quite difficult, but in parallel
illuminate the inner works of mass generation.

 On the other hand, deciding to tackle instead exclusively perturbative
problems yielding precise mathematical results would be
misleading, since exact solvability is not a physics goal {\it per
se}, whereas the correct understanding of physical phenomena is.
As reminds us Herakleitos from  6th century B.C., ``nature likes
to hide", and physics gives frequently rise to highly non-trivial
phenomena defying rigorous analytical description, quark
confinement being a well-known example.

The motivation for the analysis of the katoptron model is
basically two-fold. One comes from experimental data
\cite{george3} suggesting the existence of new particles close to
the weak scale and thus accessible to the next generation of
experiments in the not-too-distant future. The other one stems
from considerations related to a particular unification of all
particle interactions at energy scales which are high enough to
naturally suppress proton decay. This was discussed in
\cite{george2} and placed in a higher-dimensional unified
framework in \cite{geozoup}.

To prove analytically and demonstrate explicitly that the
katoptron model does indeed solve the hierarchy problem {\it via}
this unification in its essence, contrary to other currently
popular approaches, one should write down precise mathematical
formulas, which were already implied in \cite{george2} for the
symmetry breaking channel $SU(4)_{PS} \times SU(2)_{R}
\longrightarrow SU(3)_{C} \times U(1)_{Y}$, relating the various
energy scales of the model with each other. The starting point
lies in the renormalization-group equations that give rise to the
running of a given gauge coupling $g$ with momentum $p$, which is
described, to 1-loop, by the
 well-known equation
\begin{equation}
\alpha^{-1}(p) = \alpha^{-1}(p_{0}) + c\ln{(p/p_{0})},
\end{equation}

\noindent where $\alpha=\frac{g^{2}}{4\pi}$, $p_{0}$ is some
reference scale and $c = \frac{11N-2N_{f}}{6\pi}$ when the theory
contains $N_{f}$ fermions transforming under the fundamental
representation of the gauge group $SU(N)$.

 In order to derive analytical formulas for the
physical scales of the model, one has to use the following
relations having their source in unification constraints and
dynamical assumptions \cite{george2} defining in parallel the
scales $\Lambda_{PS}$, $\Lambda_{QCD}$, $\Lambda_{K}$ (denoted by
$\Lambda_{M}$ in \cite{george2}) and $\Lambda_{GUT}$ respectively:
\begin{eqnarray}
\alpha_{Y}^{-1}(\Lambda_{PS})&=&\frac{3\alpha_{L}^{-1}(\Lambda_{PS})+
2\alpha_{C}^{-1}(\Lambda_{PS})}{5} \nonumber \\
\alpha^{-1}_{C}(\Lambda_{QCD}) \sim 1, &&
\alpha^{-1}_{K}(\Lambda_{K}) \sim 1 \nonumber \\
\alpha_{L}(\Lambda_{GUT})=&\alpha_{PS}(\Lambda_{GUT})&
=\alpha_{K}(\Lambda_{GUT}) \equiv \alpha_{GUT},
\end{eqnarray}

\noindent where $\alpha_{Y,L,C,PS,K}$ are the couplings
corresponding to the gauge groups $U(1)_{Y}$, $SU(2)_{L}$,
$SU(3)_{C}$, $SU(4)_{PS}$ and $SU(3)^{\prime}$, the katoptron
generation gauge group (denoted by $SU(3)_{2G}$ in
\cite{george2}).

The definition of the following constants proves then to be
useful:

\begin{eqnarray}
A =
\alpha_{Y}^{-1}(M_{Z})&-&\frac{1}{5}\left(3\alpha_{L}^{-1}(M_{Z})
+2\alpha_{C}^{-1}(M_{Z})\right)
\nonumber \\~\\ c_{N}=\frac{11N-12}{6\pi}, &
&\tilde{c_{N}}=\frac{11N-24}{6\pi} \;\;{\rm for}\;\; N = 0,2,3,4
\nonumber
\\~\\ c_{K} = \frac{17}{6\pi},&& B =
\frac{3c_{2}+2c_{3}}{5}-c_{0} \nonumber \\~\\ E &=&
\alpha^{-1}_{C}(M_{Z})-1+(1-\alpha^{-1}_{L}(M_{Z}))\frac{\tilde{c_{4}}-c_{K}}
{\tilde{c_{2}}-c_{K}}+\frac{A(\tilde{c_{3}}-\tilde{c_{4}})}{B}
\nonumber \\~\\ F &=&
\frac{c_{2}(\tilde{c_{4}}-c_{K})}{\tilde{c_{2}}-c_{K}} +
\tilde{c_{3}}-\tilde{c_{4}}-c_{3},
\end{eqnarray}

\noindent where $A$ is expressed in terms of the values of the
gauge couplings measured at the mass $M_{Z}$ of the $Z^{0}$ gauge
boson,
 $c_{K}$ describes the running of $\alpha_{K}$, and the
values $N=0,2,3,4$ for $c_{N}, \tilde{c_{N}}$ correspond to the
couplings $\alpha_{Y,L,C,PS}$ for scales below and above
$\Lambda_{K}$ respectively \footnote{In order to derive
order-of-magnitude relations, it is assumed here that all
katoptrons decouple at scales below $\Lambda_{K}$. It will be
shown later that a relatively small mass hierarchy between
katoptrons renders this not exactly true.}.

 All the necessary ingredients are now available in order to express
 the physical scales of the model in a 1-loop approximation.
 Apart from the easily-derived $\Lambda_{QCD} =
M_{Z}\exp{\left(\frac{1-\alpha^{-1}_{C}(M_{Z})}{c_{3}}\right)}$,
assuming that $\alpha_{K}$ does not influence considerably the
rest of the gauge couplings when it becomes strong \cite{george2}
and noting  that $\tilde{c_{2}}-c_{K} \sim -1$,  one has

\begin{equation}
\Lambda_{GUT} =
\Lambda_{K}\left(\exp{(1-\alpha^{-1}_{L}(M_{Z}))}\left(\frac{M_{Z}}{\Lambda_{K}}
\right)^{c_{2}}\right)^{\frac{1}{\tilde{c_{2}}-c_{K}}} \;\sim \;
\Lambda_{K}e^{\alpha^{-1}_{L}(M_{Z}) } \nonumber \label{eq:GUT}
\end{equation}

\begin{equation}
\Lambda_{PS} = M_{Z}e^{A/B} \nonumber
\end{equation}

\begin{equation}
\Lambda_{K} = M_{Z}e^{E/F}. \nonumber
\end{equation}

In these relations, input variables are only the three coupling
constants $\alpha_{Y,L,C}(M_{Z})$ and $M_{Z}$. Using the
experimentally measured values $\alpha^{-1}_{Y}(M_{Z}) \sim 59.2$,
$\alpha^{-1}_{L}(M_{Z}) \sim 29.6$, $\alpha^{-1}_{C}(M_{Z}) \sim
8.4$ and $M_{Z} \sim 91.2$ GeV \cite{PDG}, one obtains
$\Lambda_{QCD} \sim 0.12$ GeV, $\Lambda_{GUT} \sim 5.6 \times
10^{15}$ GeV, $\Lambda_{PS} \sim 6 \times 10^{13}$ GeV, and
$\Lambda_{K} \sim 840$ GeV, which are of course consistent with
the results of \cite{george2}. If one claims further that the
value of $M_{Z}$ is determined dynamically by $\Lambda_{K}$, one
is left with only three independent input constants.

To understand how one can argue this, remember that $M_{Z}$ can be
expressed by
\begin{equation}
M_{Z} = v\sqrt{\pi(\alpha_{Y}(M_{Z})+\alpha_{L}(M_{Z}))},
\end{equation}

\noindent where $v \sim 250$ GeV denotes the weak scale. If this
scale is generated dynamically, it is approximately given by
\cite{george1}
\begin{equation}
v=
\frac{1}{2\pi}\sqrt{\sum_{i}M^{2}_{i}\ln{(\Lambda_{\chi}/M_{i})}
},
\end{equation}

\noindent where $i$ counts the new fermion electroweak doublets
introduced in the theory and $\Lambda_{\chi} \sim \Lambda_{K}$ is
the katoptron chiral symmetry breaking scale. In the following
section one can see that the contribution of the eight lighter of
the twelve new doublets of the katoptron model in this relation is
negligible, and that to avoid fine-tuning one may take the scales
$\Lambda_{\chi}$ and $M_{i}$ to be of the same order of magnitude.
In this case one has $v \;~^{<}_{\sim}\; \Lambda_{K}/\pi$, a
result consistent with the Z-boson mass, a mass which is therefore
no longer independent from $\Lambda_{K}$.

Alternatively, one could suppose that there exists a more
fundamental theory producing values for $\alpha_{GUT}$,
$\Lambda_{GUT}$ and $\Lambda_{PS}$, in which case the values of
the gauge couplings at $M_{Z}$ and the rest of the scales would be
the output of the model. Anyway, the fact that this theory allows
for the determination of unique order-of-magnitude energy-scale
relations such as the ones just given and for the transparent
solution of the hierarchy problem constitutes a quite powerful
motivation for the further study of katoptron dynamics in order to
resolve correctly the puzzle of fermion mass generation and open
the ``black box" of fermion Yukawa couplings.

\section{Gauge-symmetry breakings and the fermion masses}

\subsection{Self-breaking of $SU(3)^{\prime}$}

We start by listing the low-energy particle content of the theory
\cite{george1} since it proves useful for the discussion of this
section. Under the gauge symmetry
 $SU(3)_{C} \times SU(2)_{L} \times U(1)_{Y} \times
SU(3)^{\prime}$, fermions transform as
\begin{eqnarray}
&{\rm SM\;fermions} &{\rm Katoptrons} \nonumber
\\
&q_{L}:({\bf 3,\;2},\;1/3,\;{\bf 1})_{i}  & q_{R}^{K}:({\bf
3,\;2},\;1/3,\;{\bf 3}) \nonumber \\ &l_{L}:({\bf
1,\;2},\;-1,\;{\bf 1})_{i}  &l_{R}^{K}:({\bf 1,\;2},\;-1,\;{\bf
3})\nonumber \\&q_{R}^{c}:({\bf {\bar
3},\;1},\;~^{-4/3}_{+2/3},\;{\bf 1})_{i}  & q_{L}^{K\;c}: ({\bf
{\bar 3},\;1},\;~^{-4/3}_{+2/3},\;{\bf 3}) \nonumber
\\&l_{R}^{c}:
 ({\bf 1,\;1},\;~^{0}_{2},\;{\bf 1})_{i} & l_{L}^{K\;c}:
 ({\bf 1,\;1},\;~^{0}_{2},\;{\bf 3}),
   \nonumber \\
\end{eqnarray}
\noindent where $q$ and $l$ denote quarks and leptons respectively
and $i=1,2,3$ is a SM-generation index. It should be reminded here
that in the unified context of \cite{geozoup} this theory is
anomaly-free. The katoptron coupling $\alpha_{K}$ becomes strong
at energy scales around $\Lambda_{K}$ and the katoptrons acquire
dynamical masses in a similar way that ordinary quarks acquire
dynamical (``constituent") masses because of QCD.

Therefore, the condensate \footnote{The complex conjugate parts of
composite fermionic operators are omitted here and in the
following for simplicity. The influence of weak interactions on
the dynamics discussed is therefore neglected in this first
approach. Note also that specifying the correct fermion handedness
is indispensable for the gauge invariance of these operators.}
$<\bar{\psi_{L}^{K}}\psi_{R}^{K}>$ breaks not only the electroweak
gauge symmetry but also the katoptron chiral symmetry at the right
scale, substituting thus the elementary Higgs mechanism. In
addition, the same condensate is presumed to cause the
self-breaking of the katoptron generation group $SU(3)^{\prime}$
{\it via} the channel
 ${\bf 3} \times {\bf 3} \longrightarrow \bar{{\bf 3}}$, where ${\bf 3}$
 denotes the fundamental representation of $SU(3)^{\prime}$
  in which katoptrons reside. As is shown in the next
 subsection, the formation of this condensate proves to be very crucial
 also for the generation of SM-fermion masses.

 The channel ${\bf 3} \times \bar{{\bf 3}} \longrightarrow {\bf
 1}$ would leave the katoptron generation symmetry intact, but
 it cannot be realized with Lorenz-scalar operators (contrary to what
 happens in QCD, here both $\psi^{K}_{R}$ and $\psi^{K\;c}_{L}$ reside
 in the {\bf 3} of $SU(3)^{\prime}$). Therefore,
  assuming that gauge
 symmetries cannot break Lorenz symmetries, the katoptron group
 most likely self-breaks. Majorana masses are also not expected to
 be generated, since QCD interactions make the $SU(3)\times U(1)_{Y}$
 symmetry-preserving condensate which produces Dirac masses
  correspond to the most attractive channel.

 The self-breaking of $SU(3)^{\prime}$ indicates that
 third-generation katoptrons (denoted by a ``3" superscript below)
  acquire masses on the order of
 $\Lambda_{K}$ {\it via} the condensate
 $<\bar{\psi_{L}}^{3K}\psi_{R}^{3K}>$, justifying
  thus the approximate relation giving the weak scale
 in the previous section. The generation group is consequently
  broken down to
 $SU(2)^{\prime}$, and the corresponding coupling also becomes
 strong in its turn but at lower energies.
 To calculate the scale at which the
 new chiral symmetry breaking related to the two lighter
 katoptron generations takes place, recall that this happens when
 the relevant gauge coupling reaches the
 critical value $\alpha_{c}=\frac{\pi}{3C_{2}(R)}$ \cite{Miransky1},
  where $C_{2}(R)$ is the quadratic Casimir of the representation $R$ of the
 gauge group. For fermions transforming under the fundamental representation
 of an $SU(N)$ gauge group, $C_{2}$ is given by
 $C_{2}= \frac{N^{2}-1}{2N}$.

 Denoting the critical couplings and chiral symmetry breaking scales
 of $SU(3)^{\prime}$ and $SU(2)^{\prime}$ by $\alpha_{c}$ and
 $\tilde{\alpha_{c}}$ and by $\Lambda_{\chi}$ and
 $\tilde{\Lambda_{\chi}}$ respectively, one has
 \begin{eqnarray}
 \tilde{\alpha^{-1}_{c}} = \alpha^{-1}_{c} + \tilde{c_{K}}
 \ln{(\tilde{\Lambda_{\chi}}/\Lambda_{\chi})}
 \end{eqnarray}

\noindent where $\tilde{c_{K}} = 1/\pi$ describes the running of
the $SU(2)^{\prime}$ coupling. This relation yields
$\Lambda_{\chi} = \tilde{\Lambda_{\chi}}e^{7/4} \sim 5.75
\tilde{\Lambda_{\chi}}$, which should also express approximately
the mass hierarchy between the third and the two lighter katoptron
generations.

In view of the fact that the values of the critical couplings are
quite large, the 1-loop $\beta$-function is not very accurate and
the equation above should only be considered as a crude
approximation giving order-of-magnitude results. Furthermore, note
that the condensates $<\bar{\psi_{L}}^{(1,2)K}\psi_{R}^{(1,2)K}>$
which break the chiral symmetry of the two lighter katoptron
generations do not break the $SU(2)^{\prime}$ group. On the other
hand, QCD interactions can break the remaining katoptron
generation symmetry at lower energies by forming condensates of
the form $<\bar{q_{L}^{K}}q_{R}>$ and having $\Lambda_{QCD}^{3}$
as a natural order of magnitude.

\subsection{Mass hierarchies and mixing angles}

Having described the main qualitative features of the dynamics of
the katoptron group, we are ready to start an order-of-magnitude
calculation of the SM-fermion masses and mixing angles. To
understand why the third-generation standard-model fermions are
much heavier than the other ones, one has to study the relevant
operators in the effective Lagrangian. What is clearly needed is
the formation of multi-fermion composite operators containing
fermion bilinears $\bar{\psi}_{L}^{K}\psi_{R}$ which mix
katoptrons with SM fermions and thus provide a mass feed-down
(generalized see-saw) mechanism.

Since the original Lagrangian is chirally symmetric, the search is
focused on non-renormalizable operators arising
non-perturbatively. These might not be generated explicitly by
gauge interactions, but they should be consistent with the gauge
symmetries of the model. Unlike extended-technicolour operators in
technicolour theories, these operators are not likely to be highly
suppressed since the energy scale where the katoptron symmetry is
broken is obviously very close to the scale where the katoptron
coupling becomes strong.

Simple inspection of familiar types of operators is initially
discouraging. The operator
$\frac{1}{\Lambda_{QCD}^{2}}(\bar{\psi}_{R}
\psi_{L}^{K})\bar{\psi}_{L}^{K}\psi_{R}$ for instance is not
supported by dynamics strong enough to generate third-generation
SM fermion masses, since
$\frac{1}{\Lambda_{QCD}^{2}}<\bar{\psi}_{R}\psi_{L}^{K}>$ should
be on the order of $\Lambda_{QCD} \ll m_{t,b}$ and is more suited
for lighter SM-fermion masses. Moreover, an operator of the form
$\frac{1}{\Lambda_{K}^{2}}(\bar{\psi}_{L}^{K}\psi_{R}^{K})
\bar{\psi}_{L}^{K}\psi_{R}$ would have  strong enough dynamics,
since $\frac{1}{\Lambda_{K}^{2}}<\bar{\psi_{L}}^{K}\psi_{R}^{K}>
\sim \Lambda_{K}$, but unfortunately is not gauge invariant and
would break the electroweak symmetry explicitly.

 One is therefore lead to study higher-dimensional composite
 operators. The list of the ones quoted next is meant to be indicative
 and by no means exhaustive. Consider for instance an operator of the form
$\frac{1}{\Lambda_{K}^{5}}(\bar{\psi^{3K}_{R}}\psi^{3K}_{L})
(\bar{\psi^{3K}_{R}}\psi^{3K}_{L})\bar{\psi^{3K}_{L}}\psi^{a}_{R}$,
where $a$ stands for the various SM fermions.
 When the katoptron gauge
 coupling $\alpha_{K}$ becomes strong, katoptron condensates are
 formed and one has dynamical mass terms in the effective Lagrangian of the form
 $m_{3a}\bar{\psi}_{L}^{3K}\psi^{a}_{R}$, where
 \begin{equation}
 m_{3a}=\frac{\lambda_{3a}}{\Lambda_{K}^{5}}
 <\bar{\psi^{3K}_{R}}\psi^{3K}_{L}>^{2}
 \end{equation}

\noindent  and $\lambda_{3a}$ is an effective multi-fermion
coupling. Since $<\bar{\psi^{3K}_{R}}\psi^{3K}_{L}> \sim
\Lambda_{K}^{3}$, one finds that  $m_{3a} \sim \lambda_{3a}
\Lambda_{K}$. Note that the operator above is gauge-invariant only
when it involves katoptrons of the third generation. This makes
mass terms of the form $\bar{\psi^{K}_{L}}\psi_{R}$  much larger
for third-generation katoptrons, while for the lighter generations
these are {\it a priori} expected to be quite smaller.

This discussion renders the connection between the heaviness of
the top quark and the possibly large $\delta g_{R}^{t}$
\cite{george1} clearer. When katoptron condensates are formed, the
non-perturbative operator $\frac{1}{\Lambda_{K}^{8}}
(\bar{\psi^{3K}_{L}}\psi^{3K}_{R})(\bar{\psi^{3K}_{L}}\psi^{3K}_{R})
\bar{\psi^{3K}_{L}}\psi^{a}_{R} \bar{\psi^{3K}_{L}}\psi^{a}_{R}$
(the lowest-dimensional gauge-invariant operator relevant to
$\delta g_{R}^{t}$ which involves katoptron dynamics) becomes
proportional to
$\frac{m_{3a}}{\Lambda_{K}^{3}}\bar{\psi^{3K}_{L}}\psi^{a}_{R}
\bar{\psi^{3K}_{L}}\psi^{a}_{R}$. However, this operator is
conjectured to  be responsible for the deviation of the weak
couplings $g_{R}^{t,b}$ from their standard-model values and
consequently for the smallness of the $S$ parameter
\cite{george1}. Since the mass relation $m_{t^{K}} \sim m_{b^{K}}$
is expected to hold, the fact that $m_{t} \gg m_{b}$ translates
into $m_{3t} \gg m_{3b}$ within the generalized see-saw framework
of the model, and explains why the anomalous-coupling relation
$\delta g_{R}^{t} \gg \delta g_{R}^{b}$ is plausible.

A similar higher-dimensional operator could make
electroweak-invariant mass terms not involving third-generation
katoptrons larger than the naively supposed $\Lambda_{QCD}$ scale.
Consider for instance the operator $\frac{1}{\Lambda^{8}}
(\bar{\psi^{3K}_{L}}\psi^{3K}_{R})(\bar{\psi^{3K}_{R}}\psi^{3K}_{L})
(\bar{\psi^{2K}_{L}}\psi^{a}_{R})
\bar{\psi^{2K}_{L}}\psi^{a}_{R}$, where $\Lambda$ is some relevant
energy scale. At low-enough energies, we have the formation of
condensates due not only to QCD but to katoptron interactions as
well. The operator then becomes $\frac{1}{\Lambda^{8}}
<\bar{\psi^{3K}_{L}}\psi^{3K}_{R}><\bar{\psi^{3K}_{R}}\psi^{3K}_{L}>
<\bar{\psi^{2K}_{L}}\psi^{a}_{R}> \bar{\psi^{2K}_{L}}\psi^{a}_{R}
\sim m_{2a}\bar{\psi^{2K}_{L}}\psi^{a}_{R}$ with $m_{2a} =
\lambda_{a}\Lambda_{QCD}^{1-\epsilon_{a}}\Lambda_{K}^{\epsilon_{a}}$,
where the parameters $\lambda_{a}$ and $\epsilon_{a}$ should be
determined by the non-perturbative dynamics of the model.

 It is apparent that these mass terms could be as large as
 $\Lambda_{K}$ according to the values
assumed by $\epsilon_{a}$, the computation of which lies however
beyond the scope of this letter. It might just be added that in
principle $\epsilon_{a}$ is momentum-dependent, and a simple
relevant ansatz would be $\epsilon_{a}=\gamma_{a}\ln{(p_{0}/p)}$,
with $\gamma_{a}$ an appropriate positive anomalous dimension. The
appearance of such terms is not surprising, since no symmetry
protects these masses from being large after the katoptron
generation group is broken. The fact that the $SU(2){\prime}$
coupling is already strong just before this symmetry is broken
could accentuate this effect. Unfortunately, nature has not given
us yet examples of gauge symmetries broken below the scale where
their couplings become strong and the relevant dynamics are hard
to pin down. It is assumed next that $\epsilon_{a}$ is
particularly large for third-generation SM fermions, something
that proves to be convenient for the numerical exercise below.

The mass-matrix example given in \cite{george1} can now be
improved by considering merely for illustration purposes the
following mass matrices, which for simplicity are taken to be real
and have the form
\begin{eqnarray}
&& {\cal M}_{i} = \left(\begin{array}{cc}  0 & m_{i} \\ m_{i} &
M_{i}
 \end{array} \right), i = U,D
\end{eqnarray}

\noindent for the up-type ($U$)and down-type quarks ($D$), with
$M_{i},m_{i}$  symmetric (similar matrices can be constructed for
leptons \cite{george2}):

\begin{eqnarray}
\begin{array}{c}  \\ m_{U} ({\rm GeV}) =\\ \end{array} &&
\left(\begin{array}{ccc} 0.7 & 0.8 & 15 \\ 0.8 & 0.8 & 73 \\ 15 &
73 & 410 \end{array} \right)
\begin{array}{c}  \\,\;\;  m_{D} ({\rm GeV}) =\\ \end{array}
\left(\begin{array}{ccc} 0.7 & 0.8 & 1 \\ 0.8 & 0.8 & 15 \\ 1 & 15
& 45 \end{array} \right).
\end{eqnarray}

\noindent The dynamical assumption is made again that the
$SU(2)_{L}\times U(1)_{Y}$-breaking katoptron dynamical mass
submatrices are diagonal and have the form
\begin{eqnarray}
\begin{array}{c}  \\ M_{U}=M_{D} ({\rm GeV}) =\\ \end{array} &&
\left(\begin{array}{ccc} 170 & 0 & 0 \\ 0 & 170 & 0 \\ 0 & 0 &
1000
\end{array} \right).
\end{eqnarray}

Note that the matrices $M_{U,D}$ are taken to be equal to each
other in order to respect isospin symmetry and avoid large
contributions to the $T$ parameter, since the $m_{t}-m_{b}$
hierarchy is reproduced by terms in the $m_{U,D}$ matrices which
are electroweak singlets. The hierarchy between the third and the
two lighter katoptron generations is consistent with our previous
discussion and interestingly enough proves to be crucial for the
correct reproduction of the SM-CKM matrix. Moreover, the terms in
$m_{U,D}$ not involving third-generation fermions do not exceed
the GeV scale, in accordance with the previous considerations.

The rigorous study of the dynamics producing these specific mass
entries corresponding to particular values of $\epsilon_{a}$ and
$\lambda_{a}$ introduced before, and which give rise to the large
mass hierarchy between up- and down-type quarks or for instance
between the quarks and leptons in the SM requires a lengthier
investigation. It is interesting to note however that
near-critical interactions in dynamical-symmetry-breaking theories
may typically reproduce such hierarchies \cite{Miransky2}. The
considerations in that reference are readily applicable here as
well, since the electroweak-invariant mass terms that mix
SM-fermions with their partners and give them mass are generated
by critical dynamics which are in addition responsible for the
katoptron-symmetry-breaking condensate.

 After diagonalisation of the mass matrices,
 the quark masses are found to be approximately equal (in GeV units and
 renormalized at TeV scales) to
\begin{eqnarray}
165, 0.75, 0.001 \;\;\;\;\;{\rm \;Up-type \;SM \;quarks}\\ 170,
184, 1152 \;\;\;\;\;{\rm \;Up-type \;katoptron \;quarks}\\ 3.5,
0.06, 0.003\;\;\;\;\;{\rm \;Down-type \;SM \;quarks}
\\ 170, 171, 1002\;\;\;\;\;{\rm \;Down-type \;katoptron \;quarks}
\end{eqnarray}

\noindent which, along with the corresponding leptons, reproduce a
correct order of magnitude for the weak scale. Therefore, as
regards particles carrying QCD color, this model predicts the
existence of four new fermions not much heavier than the top
quark, and two more with masses around 1 TeV.

The mass matrices introduced give rise to a unitary generalized
CKM matrix describing the mixing between the fermions of the
theory, with a non-unitary submatrix $V_{SM-CKM}$ corresponding to
the standard-model fermions given (in absolute values) by
\begin{eqnarray}
\begin{array}{c}  \\ | V_{SM-CKM} | \sim \\ \end{array} &&
\left(\begin{array}{ccc} 0.97 & 0.23 & 0.008 \\ 0.22 & 0.97 & 0.07
\\ 0.005 & 0.06 & 0.95 \end{array} \right),
\end{eqnarray}

\noindent which is reasonably close to the experimentally measured
SM-CKM matrix (obviously not assuming SM-CKM unitarity and taking
renormalization to higher scales into account). Note the small
predicted value of $|V_{tb}| \sim 0.95$ which is due to the large
mixing of the top quark with its katoptron partner and is
obviously related to the large value of $m_{t}$.

\section{Conclusions}

Previous attempts to introduce katoptrons in physics beyond the SM
had already made clear how electroweak symmetry can be broken
dynamically by non-perturbative effects. However, the source of
SM-fermion masses was still left rather obscure. The present work
constitutes a first attempt to clarify certain qualitative aspects
of fermion mass generation in connection with the katoptron
gauge-group self-breaking and to identify the reasons for the
appearance of various mass hierarchies with no recourse to
unnaturally light elementary scalar particles. A detailed
computation of fermion masses within this framework would entail a
big effort tackling strong-dynamics issues, but in parallel would
complete the picture of katoptron theory.

Moreover, apart from its theoretical consistency the model
possesses clear experimental signatures which place it on solid
epistemological grounds and deserve to be looked at in future
experimental projects. \footnote{ A section of a letter written by
the philosopher Epicure to Pythocles more than two millennia ago
is enlightening at this point: ``But when one accepts one theory
and rejects another which harmonizes just as well with the
phenomenon, it is obvious that one altogether leaves the path of
scientific enquiry and has recourse to myth"}.
 If the scheme proposed finally proves to be true, it would be the
first known case in nature of a gauge symmetry breaking itself and
another gauge symmetry due to its non-perturbative dynamics. Mass
as a physical quantity would therefore have its source in gauge
interactions; and the mass of the presently known elementary
particles would be a manifestation of mixing with their - now
merely virtual - katoptron partners, since these must have decayed
just after the creation of our world.

\noindent {\bf Acknowledgements} \\ The author thanks the Physics
Department of NTUA for their hospitality and G. Zoupanos for
interesting discussions and comments on the manuscript.

\end{document}